\documentclass[%
 reprint,
 amsmath,amssymb,
 aps,
floatfix,
]{revtex4-2}
\usepackage{graphicx}
\usepackage{dcolumn}
\usepackage{bm}

\begin{document}

\preprint{APS/123-QED}

\title{Resonant phase matching \\ by oblique illumination of a dielectric laser accelerator}

\author{S. Crisp}
\author{A. Ody}
\author{P. Musumeci}
\affiliation{UCLA Department of Physics and Astronomy, 475 Portola Plaza, Los Angeles, CA, 90095}
\author{R. J. England}
\affiliation{SLAC National Accelerator Laboratory, 2575 Sand Hill Rd, Menlo Park, CA, 94025}

\date{\today}

\begin{abstract}
In dielectric laser-driven accelerators (DLA), careful tuning of drive-laser wavelength and structure periodicity is typically required in order to hit the resonant condition and match the phase velocity of the accelerating wave to the electron beam velocity. By aggressively detuning (up to 30 mrad) the angle of incidence of the drive laser on a double grating DLA structure, we show that it is possible to recover resonant phase matching and maximize the energy modulation of an externally injected 6 MeV beam in a 800 nm period structure driven using a 780 nm laser. These results show that it is possible to power DLA structures away from their design working point, and excite accelerating fields in the gap with phase profiles that change by a relatively large amount period-to-period. This flexibility is a key feature of DLAs and a critical element in the realization of phase-modulation based ponderomotive focusing to demonstrate MeV energy gain and large capture in a single DLA stage. 
\end{abstract}

\maketitle


\section{\label{sec:Intro}Introduction}
Dielectric laser accelerators (DLA) hold the promise to shrink the size of relativistic particle accelerators by 4 orders of magnitude due to their micron-size characteristic scales, while at the same time increasing the accelerating gradient by more than one order of magnitude \cite{peralta:dla, breuer:DLA, Cesar2018, Leedle:DLA}. Unique advantages of DLAs compared to other laser accelerators include use of modest laser powers and the possibility for very high efficiency acceleration \cite{siemann,England2014_DLARev}. Modern nanofabrication techniques enable on-chip manufacturing for extreme miniaturization of the devices \cite{Sapra2020}, opening interesting prospects for a variety of applications including linear colliders, radiation generation, and medical use \cite{Wootton2016App}. 

One important characteristic of DLAs is the temporal format of the output beams which are microbunched on the scale of the laser wavelength \cite{black2019, schonenberger2019}. While optically microbunched beams have been demonstrated in various setups ranging from high energy particle accelerator beamlines \cite{Sears2008Microbunch, sudar:doublebuncher} to electron microscope columns \cite{FeistRopers2017}, DLAs offer unique advantages towards solving the problem of coupling infrared and visible light to moderately relativistic electrons. In fact, at higher energies ($>$ 50 MeV), the FEL interaction is very effective in microbunching the electrons at the attosecond scale \cite{Duris2020_XLeap,Sears2008Microbunch}. At lower energies ($<200$ keV), small nanostructures can be used to simply couple the light to the electrons \cite{KealhoferBaum2016}. In the intermediate energy range (1-10 MeV), uniquely useful for ultrafast electron diffraction and scattering applications \cite{hawkes:springerhandbook}, the demonstration of attosecond bunch trains is still far from having been realized. 

One of the challenges in DLA is the relatively small size of the stable accelerating region in the longitudinal phase space. The relevant parameter setting the energy acceptance of the accelerator is the normalized wave amplitude $\alpha = e_0 E_0 / m_e c^2 k$ where $k = \omega /c$ is the laser free-space wavenumber. The relative energy acceptance is proportional to the square root of this quantity, which is typically very small ($10^{-5}$) for DLA, especially compared to $\alpha \approx 1$ in conventional RF accelerators. Consequently, one has to carefully design DLA structures to be resonant (and to efficiently interact) with the target energy electron beam. As an example, in order to excite waves phase-synchronous with 6 MeV electrons, an 800 nm period structure needs to be driven by an 803 nm laser. If the laser is detuned to 804 nm, the dephasing length (the length of the interaction region over which the accelerating wave and the electrons remain synchronous so that effective interaction can take place) reduces to $<$ 300 $\mu$m strongly limiting the achievable energy gain. 

At the same time, there are a variety of instances where it might be convenient to drive a DLA structure using a laser source at a different frequency than the design value, or, equivalently, an input e-beam energy different than originally planned. This is also relevant in setups where the spectrum of the drive laser might not be perfectly monochromatic or is not precisely centered around the design wavelength. In some instances, in order to control the transverse dynamics, a quickly varying phase profile might be added to the laser (therefore broadening the incoming spectrum), as recently proposed \cite{cesar:alloptical}. From the fabrication point of view, manufacturing limitations often constrain the physical dimensions of the DLA features to non-optimized (typically larger) sizes so that driving a structure with a different laser wavelength might even yield improvements in the diffraction efficiency. All these cases indicate that the possibility to compensate in-situ for structure mismatch could be an important factor in large scale DLA development.

\begin{figure*}[!ht]
    \centering
    \includegraphics[width = \textwidth]{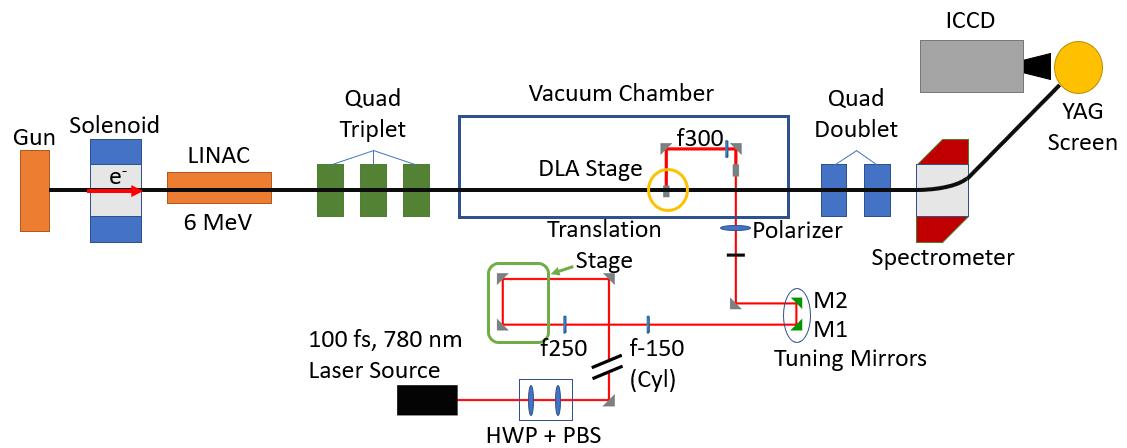}
    \caption {Schematic of the Pegasus beamline for DLA acceleration. A 1-2 pC beam generated by a modified SLAC/UCLA/BNL RF photogun \cite{alesini} is accelerated by the linear accelerator cavity (LINAC) which is tuned to reduce the energy spread to below 30~keV full width half max. The gun solenoid and a quadrupole (quad) lens triplet are used to focus the beam through the central vacuum chamber, where the DLA sample is located. After the chamber, a quad doublet is used to refocus the electrons onto a YAG screen located on the high resolution dipole spectrometer. This final screen is viewed with an intensified camera. The optical system for the drive laser includes 2 achromatic lenses and a cylindrical lens to shape the intensity distribution at the DLA. A half wave plate (HWP) and polarising beam splitter (PBS) combination allow intensity control. In addition, there is a translation stage for timing alignment and tunable mirrors (M1, M2) for angle of incidence control.}
    \label{fig:beamlinescheme}
\end{figure*}

In this paper, we demonstrate the flexibility for double grating DLA structures \cite{Peralta2012Structure, wei2017dual} to be driven using a laser system of wavelength shorter than the structure periodicity. If illuminated at normal incidence, the phase velocity in these structures would be larger than $c$ and not suitable for particle acceleration, but by varying the incident angle from the normal by as much as 30~mrad, resonance with relativistic electrons is recovered and the energy modulation maximized. These results provide an experimental validation of the wide acceptance bandwidth of DLA structures, confirming that is possible to have large phase shifts in adjacent periods of the DLA \cite{naranjo:ponderomotive,niedermayer:APF}. The latter is a critical aspect for the next generation DLA experiments where long interaction regions and detailed control of the dynamics using external drive laser phase manipulation will be implemented \cite{Cesar:PFT, Ody:SHARD}. 

The paper is organized as follows. In the next section, the experimental setup is discussed. A sophisticated analysis procedure is developed to extract from the measured energy spectra the important parameters for the interaction, such as the accelerated charge and the induced energy modulation. We then report on measurements of the energy modulation experienced by the particles as a function of the relative laser-electron timing, and of the incident laser power to establish the spatio-temporal overlap and characterize the structure response. Finally, we present the measurements of energy modulation as a function of the incidence angle. The significance of these results is discussed before drawing general conclusions.


\section{Experimental Setup}\label{sec:exp}

The experiment was performed at the UCLA Pegasus advanced RF photoinjector beamline \cite{alesini, maxson}. The laser drive for this experiment is a (up to 30 mJ per pulse) Ti:Sapphire laser with 10 Hz repetition rate, a center wavelength $\lambda_l$ of 780 nm, and a pulse length of 100 fs FWHM. The laser pulse is split to be used both to drive the copper photocathode in the SLAC/UCLA/BNL RF gun after a third harmonic generation stage and to power the DLA structure. The 1-2~pC electron beam generated in the gun is then focused by the gun solenoid into linac to accelerate up to 6~MeV. A series of quadrupole lenses arranged in a triplet configuration is used to focus the electrons into the DLA gap. A diagnostic section comprised of two quadrupoles and a spectrometer dipole is used to monitor the beam energy distribution after the DLA interaction. The DLA-driver laser arm delivers up to 1~mJ of energy to the DLA sample. The 12 m long transport contains a series of lenses to adjust the spot size at the interaction, a half-waveplate polarizing beam splitter combo for intensity control, a translation stage for timing alignment, and the vacuum window as described in Fig. \ref{fig:beamlinescheme}.

The system parameters are reported in Table \ref{tab:beam_params}. The DLA structure for this experiment is made by separate gratings having period $\lambda_g$ = 800~nm etched on two SiO2 wafers, which are subsequently bonded together \cite{peralta:dla, Peralta2012Structure}. The structure employed here was used previously at UCLA to demonstrate record acceleration gradients \cite{Cesar2018} and energy gains \cite{Cesar:PFT}.

\begin{table}[!hb]
    \centering
    \begin{tabular}{l|l}
         \textbf{Parameter} & \textbf{Value} \\ \cline{1-2}
         Beam Energy & 6 MeV \\ 
         Beam Energy Spread (FWHM) & 23.5 keV \\
         Beam Charge & 1-2 pc \\
         Beam RMS size at DLA & 50 $\mu$m \\
         Beam Length & .25 ps \\
         Laser pulse length (FWHM) & 100 fs \\
         Laser spot size at DLA (FWHM) & 1.5mm x 345 $\mu$m\\
         Laser Energy & 1 mJ \\
         DLA length & 500 $\mu$m\\
         DLA vacuum gap & 800 nm\\
    \end{tabular}
    \caption{List of parameters for the DLA experiment. Laser spot size varied; parameters given for drive laser intensity scan.}
    \label{tab:beam_params}
\end{table}

\begin{figure}
    \centering
    \includegraphics[width = .45\textwidth]{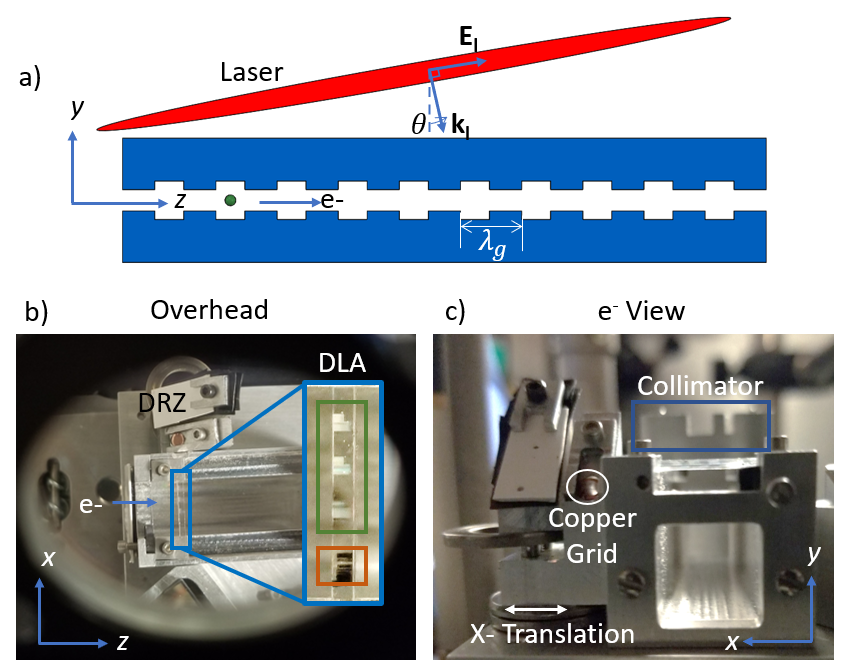}
    \caption{a) Schematic of the double grating DLA concept. A laser pulse is illuminating at angle $\theta$ with respect to the y-axis the 500 $\mu$m long DLA structure. The electrons travel along $z$ in a 800 nm wide gap. b) and c) Top and side views of the sample holder used in the experiments. The sample holder is designed to serve multiple functions. A standard TEM copper grid is used to establish electron laser synchronization \cite{Scoby2013}. A featured collimator piece provides a large reference for the position of the sample holder. Outlined in red, narrow alignment channels in the DLA structure can be used to fine-tune the beam location. A DRZ screen is also mounted on the sample holder to check the spatial overlap with the laser using an overhead camera. After insertion of the DLA sample, a signature of where the electron beam hits the sample (from the radiation emitted by the non-transmitted electrons passing in the fused silica) can also be picked up by this camera enabling a final 'live' spatial alignment in grating structure, outlined in green (3 grating sections).}
    \label{fig:DLA_Stage}
\end{figure}

Details of the sample area geometry are shown in Fig. \ref{fig:DLA_Stage}. A featured collimator is placed after the sample to filter electrons which do not pass through the 800~nm sample gap. The top of the collimator part is manufactured with a crenel-like shape that facilitates the initial alignment phase in the experiment. When the electron beam is over focused the features of the collimator create a shadow map that allows identification of the reference features on the sample. To refine the alignment, two relatively large (250 $\mu$m x 500 $\mu$m ) alignment channels etched in the SiO2 DLA substrate enable final tuning of the sample pitch and yaw angles to within 0.5 mrad. Because the beam size is approximately 50 $\mu$m RMS, and the gap in the sample is only 800~nm with an angular acceptance of 0.8 mrad, careful sample alignment is imperative to ensure that enough electrons are propagated to the spectrometer. Less than .1\% of the electrons incident on the DLA are transmitted, corresponding to up to 5000 e- per shot. The collimator helps to distinguish the transmitted electrons from the electrons passing around the sample. Gating the ICCD camera looking at the YaG spectrometer screen allows to further remove the dark current background. 

The final spectrometer screen (YAG in Fig. \ref{fig:beamlinescheme}) is energy-calibrated by measuring the change in beam position at the screen when changing the spectrometer dipole current. This yields a 0.9 keV/pixel calibration, under the assumption of a linear relation between dipole current and magnetic field. As an independent check, we also calculate the dispersion function at the screen to be 31.6 cm using the dipole radius of curvature, the bending angle and the screen location. Taking into account the spatial calibration of the camera (41 $\mu$m/pixel), we obtain a 0.78 keV/pixel calibration, indicating a slight saturation of the dipole. This latter (more conservative) value is used to convert spread in pixel into energy modulation in the rest of the paper. The point spread function of the screen and optical system allow to resolve a minimum line-width of 3 pixels corresponding to 2.34 keV.

\begin{figure}
    \centering
    \includegraphics[width = .45\textwidth]{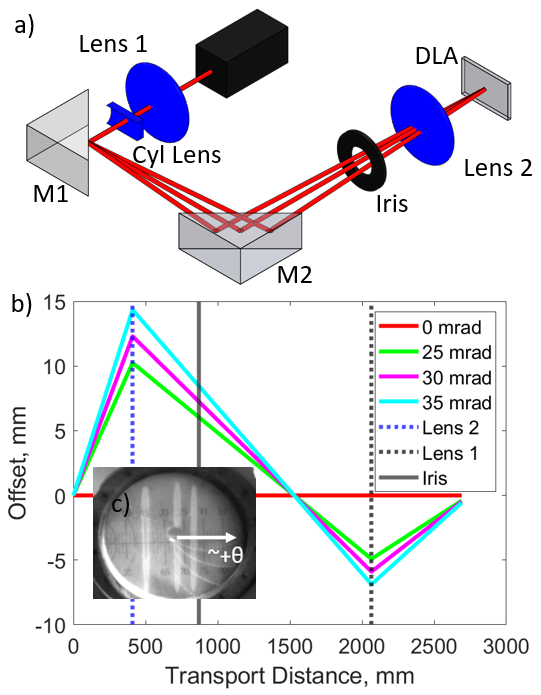}
    \caption{a) Schematic of the optical system delivering the laser to the DLA. Two achromatic lenses are used to obtain a small round spot size (FWHM = 230 $\mu$m) at the DLA plane. A cylindrical lens is then added to the transport in order to expand the laser beam along the acceleration axis. The angle of incidence is controlled by two mirrors before the reference iris, also shown. The displacement of the laser on that iris is converted to an angular offset, $\theta$, at the DLA grating. b) Reverse ray-tracing with three example trajectories at 25, 30, and 35 mrad incidence angles. The dotted lines indicate the 2 achromatic lenses, and the solid vertical line is the iris position. c) Three photos of the reference iris at different incidence angles are overlaid. There is a linear relationship between the angle of incidence at the sample $\theta$ and the displacement at the iris.}
    \label{fig:angle_schematic}
\end{figure}

The temporal synchronization of the laser and electron pulses is initially established by focusing the laser on a standard transmission electron microscopy (TEM) copper grid mounted at a $45^{\circ}$ angle next to the DLA structure with a fluence of $1 J/cm^2$, acquired by removing the cylindrical lens from the transport. The fields of the electron cloud liberated at the copper surface by the laser are sampled by the passing relativistic e-beam and temporal overlap can be detected by observing their effect on the beam profile on a downstream YAG screen. A translation stage on the optical table allows us to scan the relative time-of-arrival with a clearly discernible effect lasting about 3 picoseconds \cite{Scoby2013}.

The angle of incidence of the laser on the sample can be tuned via two mirrors outside of the vacuum box. Figure \ref{fig:angle_schematic} shows a visual of this system. The laser is initially set to $\theta = $30~mrad angle of incidence onto the DLA sample by using a green alignment laser to observe the back reflection from the fused silica. This 30~mrad angle phase-matches ultrarelativistic electrons and laser fields in the structure, as calculated in Section \ref{sec:measurements}. An iris directly before the box is then used to mark this initial position. An optical ray tracing transport calculation yields the relationship between position offset at the iris and angular offset at the sample. If the position of the laser on the sample is kept constant (as monitored by the overhead camera), a 5~mm offset at the iris corresponds to a 20.7~mrad change to the angle of incidence. In order to tune the incidence angle we first tilt mirror M1 to the desired position on the reference iris, and then mirror M2 is tilted back to keep the position on the sample fixed. 

 
\section{Analysis of spectrometer images to retrieve maximum energy modulation}\label{sec:Analysis}

\begin{figure}
    \centering
    \includegraphics[width = .45\textwidth]{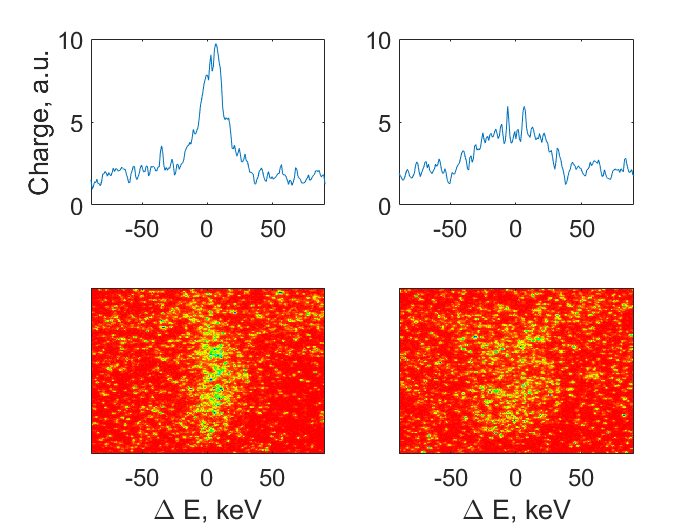}
    \caption{Two sample ICCD spectrometer images (bottom) with their corresponding lineouts (top) are shown. On the left, the unmodulated beam has a Lorenzian shape with a FWHM of 23.5 keV. A modulated spectra is shown in the right column.}
    \label{fig:spectrometerEx}
\end{figure}

The laser-on spectrometer images are analyzed to extract the induced energy modulation by the DLA process. Two sample images and the retrieved energy spectra with and without laser are shown in Fig.\ref{fig:spectrometerEx}. For monoenergetic electrons injected at random phases into the DLA, the cumulative distribution function of the energy deviation ($\Delta E$) after the accelerator can be written as
\begin{equation}\label{eq:f1}
    f_0(\Delta E) = \frac{1}{\pi}\frac{1}{\sqrt{\Delta\xi^2-\Delta E^2}},
\end{equation}
a so-called arcsin distribution sharply peaked at $\Delta E = \pm \Delta \xi$ which represents the max energy modulation of a particle in the structure and is proportional to the laser field amplitude and the length of the interaction. 

\begin{figure*}
    \centering
    \includegraphics[width =\textwidth]{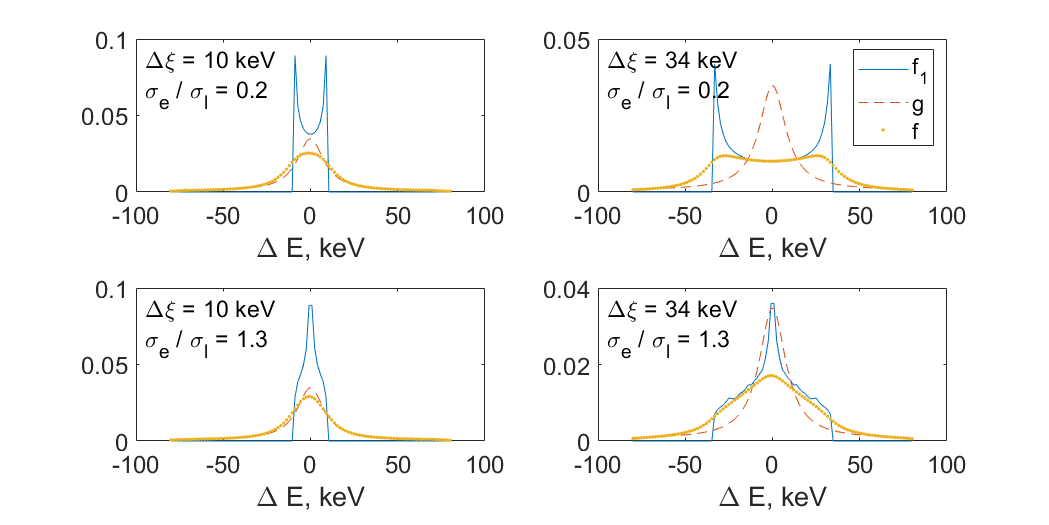}
    \caption{Simulation of the energy profiles at the spectrometer (yellow lines) when the finite electron and laser spatial distributions are taken into account. The $f$ traces are obtained by convolving the intrinsic energy distribution function $g$ (red dashed) with the induced energy modulation $f_1$ (blue dotted). The top traces correspond to a small $\sigma_e/\sigma_l$ ratio and a nearly 1D interaction. For a relatively small $\Delta \xi$ (left column), the DLA interaction only broadens slightly the initial energy spectrum (shown in red). When the DLA peak energy modulation $\Delta \xi$ is much larger than the spectrometer intrinsic resolution (right column), the characteristic double peak electron distribution is obtained. The larger $\sigma_e/\sigma_l$ ratio case is shown in the bottom row. In this case, which is the more typical case in our experimental setup, the electron beam experiences a wide range of laser intensities and the final energy profile is significantly different than the 1D case.}
    \label{fig:convEx}
\end{figure*}

In absence of any interaction, at the spectrometer the beam shows an intrinsic beam energy spectrum (which depends on the photoemission process as well as the gun and linac settings and the spectrometer resolution) with functional dependence $g(\Delta E)$. In our case $g$ is obtained by fitting the observed energy lineout without any laser modulation to a Lorentzian distribution
\begin{equation}
    g(\Delta E) = \frac{1}{\pi}\frac{\frac{\Gamma}{2}}{(\Delta E)^2 + (\frac{\Gamma}{2})^2}
    \label{eq:g}
\end{equation}
having energy width $\Gamma$ = 23.5~keV. 

When the drive laser is turned on, the observed lineout becomes the convolution of these distributions
\begin{equation}\label{eq:f3}
    f(\Delta E) = f_0 \left(\Delta E \right) \circledast g(\Delta E).
\end{equation}
A numerical deconvolution can be performed to retrieve $f_0$, but due to the noise in the measurement, the peak induced energy modulation $\Delta\xi$ is not always immediately extractable from the deconvolved profile. 

So far we assumed that all electrons see the same field amplitude in the DLA structure. Even though in general the setup is arranged to maximize the number of electrons sampling the peak of the accelerating wave, in practice this is limited by finite transverse  sizes of the beams in the interaction. In what follows we present an analysis that takes into account the influence of the transverse sizes of the electron and laser beams, $\sigma_e$ and $\sigma_l$ respectively on the energy modulation measurement.

The analysis starts by including the effect of the spatial distributions of the laser and electron beams on the final energy profile. Assuming ballistic trajectories so that electrons have constant horizontal offset $x$ when passing through the DLA structure, the energy modulation in the DLA can be written $\Delta\xi e^{-\frac{x^2}{2\sigma_l^2}}$ where the laser field is assumed to have a gaussian transverse distribution of width $\sigma_l$. The total modulation function $f_1(\Delta E)$ is given by an integral over the different electric fields experienced by the particles weighted by the electron transverse distribution function
\begin{equation}
    f_1(\Delta E, \sigma_e, \sigma_l) = \frac{1}{\pi}\int{ \frac{\frac{1}{\sqrt{2 \pi}\sigma_e} e^{-\frac{x^2}{2\sigma_e^2}}} {\sqrt{\left(\Delta\xi e^{-\frac{1}{2}\left(\frac{x}{\sigma_l}\right)^2}\right)^2-\Delta E^2}}dx}.
\end{equation}
which can be considered a 2-d generalization of $f_0$. The final modulation distribution is obtained convolving as before  $f_1$ with the initial $g(\Delta E)$ to allow a quantitative comparison with the measured electron spectra. In Fig. \ref{fig:convEx} we show two representative cases where the peak energy modulation $\Delta \xi$ is respectively comparable and much larger than the spectrometer energy resolution. When the laser beam is much wider than the electron beam (spot size ratio $\sigma_e/\sigma_l =  0.2$), the induced modulation functions $f_1$ are similar to the 1D case $f_0$. However, for beams of comparable size, the characteristic double-horn shape in the energy profile disappears and a smoother profile obtained.

\begin{figure}
    \centering
    \includegraphics[width = 0.45\textwidth]{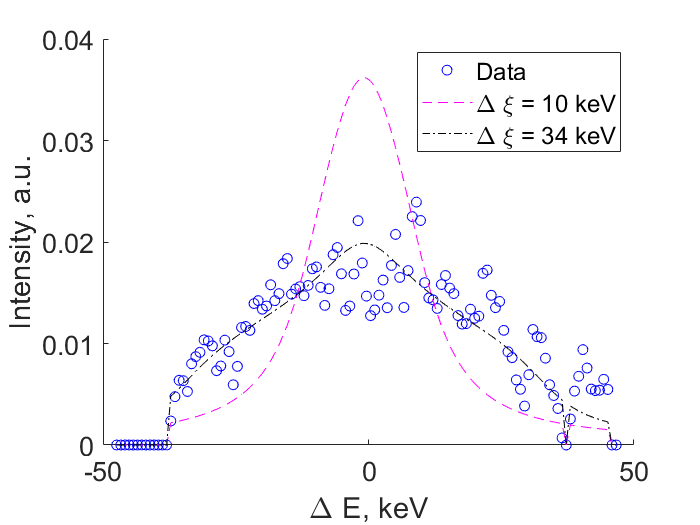}
    \caption{Example of peak energy modulation fit for the measured energy lineout (blue points), as seen in the spectrometer image on the right of Fig. \ref{fig:spectrometerEx}. In this case the fit yields a spot size ratio $\sigma_e/\sigma_L = 1.3$, and a peak energy modulation $\Delta \xi$ = 34 keV. To give an idea of the sensitivity of the fit, we show what the curve would look like for a different energy modulation $\Delta \xi =$ 10 keV for which the RMSE with the data is more than tripled.}
    \label{fig:fitComp}
\end{figure}

These simulated distributions can be directly compared to the experimental output using any number of statistical tools to retrieve the relative transverse size, $\sigma_e/\sigma_l$, and the magnitude of the maximum modulation $\Delta \xi$. Here, we use the root mean square error (RMSE) as a figure of merit for the difference between the simulated and observed spectra. All the simulated and measured energy profiles are normalized to 1, and the tails of the experimental spectra are set to zero to eliminate the edge effects of image noise. An example of a fit is shown in Fig. \ref{fig:fitComp}. This process can then be repeated for each experimental spectrum to identify the interaction parameters. We observe that the transverse size of the e-beam and the laser might vary from day to day, but are consistent in each run day. In order to take this into account we perform a global fit to retrieve the ratio $\sigma_e/\sigma_l$ which minimizes the total RMSE for each day's data set. For example, the analysis of the RMSE for the laser intensity scan data set is formed by 27 experimental lineouts indicating $\sigma_e/\sigma_l = 1.3$ during that particular run, which is then used to extract $\Delta \xi$, as in Fig. \ref{fig:fitComp}. 

\section{Measurements and Discussion}
\label{sec:measurements}

\begin{figure}
    \centering
    \includegraphics[width = 0.45\textwidth]{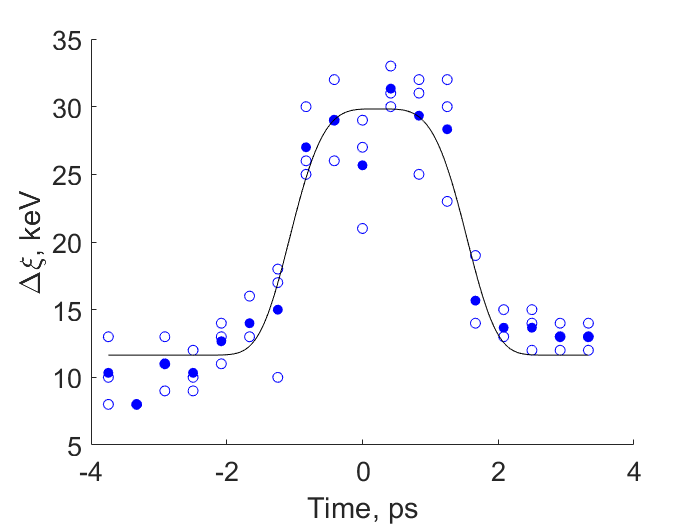}
    \caption{DLA peak energy modulation vs. relative time of arrival of laser and e-beam at the DLA. The solid circles are the mean of the three highest modulations at each delay (hollow circles). The black curve is a supergaussian fit which is used to extract the 2.6~ps FWHM for the temporal window of the interaction.}
    \label{fig:timing}
\end{figure}

Measuring $\Delta \xi$ as a function of the relative time-of-arrival of the laser and the electron beam on the DLA structure, we obtain the curve shown in Fig. \ref{fig:timing}. Strong DLA interaction with an energy modulation up to 30~keV is observed for a 2.6~ps time window, in agreement with the convolution of the 0.25~ps electron bunch duration and time-of-flight of the electrons through the DLA grating. Even after the deconvolution, the intrinsic spread of the beam and the statistical noise associated with the low electron counts contribute to a  minimum detectable floor for the measurement of $\Delta \xi$ of 12 keV.


\begin{figure}
    \centering
    \includegraphics[width = .45\textwidth]{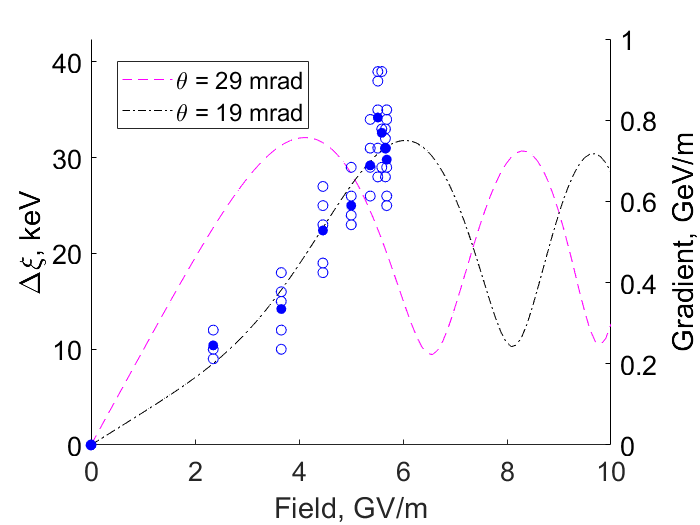}
    \caption{Observed gradient vs. incident electric field. The solid circles are the average of the data (hollow circles) for each laser energy setpoint in the scan. The pink dashed curve shows the predicted energy modulation at the phase-matching condition. A small change in the angle of incidence on the structure (10 mrad offset) yields the black line which more closely matches the measured behavior. The saturation point is reached at 6 GV/m, with a corresponding peak acceleration gradient of 0.8 GeV/m. Damage threshold and available laser energy limited us in probing the structure at higher fields.}
    \label{fig:gradVfield}
\end{figure}

In a second measurement scan reported in Fig. \ref{fig:gradVfield}, we study how the peak induced energy modulation, $\Delta \xi$, is dependent on the drive laser intensity in order to characterize the double grating DLA and extract the structure factor, $\kappa$, defined as the ratio between the peak gradient and the incident electric field. The laser energy incident on the DLA is controlled by a half waveplate and polarizing beam splitter placed in the transport line before the vacuum window.

The peak accelerating gradient is plotted on the right y-axis in Fig. \ref{fig:gradVfield}. The values are obtained from the measured peak energy modulation $\Delta \xi$ divided by the effective length, $L_{eff}$ of the interaction. In our case, with no pulse front tilt applied and a side-illumination geometry, $L_{eff}$ can be estimated as the laser amplitude (not intensity) pulse length i.e. $L_{eff} = \sqrt{2} c \tau_l$, where $\tau_l = 100$ fs.

In order to better understand the data, we use the ballistic approximation to calculate the peak energy gain of an optimally phased particle in the DLA as \cite{Cesar2018}
\begin{equation}\label{eq:nonlin}
    \Delta \xi = q \kappa \left \lvert \int_{-L/(2c)}^{L/(2c)}{E_0(t) e^{i \Delta \phi_{PM}(z(t),\theta,t) + i \Delta\phi_{NL}(I_0,t)} c dt} \right \rvert
\end{equation}
where $E_0(t)$ is the laser pulse field amplitude and $L$ is the structure length. The integration is performed along the particle trajectory inside the DLA structure. The two complex exponential phase terms $\Delta\phi_{PM}$ and $\Delta\phi_{NL}$ refer to the contributions due to phase matching and nonlinear effects, respectively. Some care is required to retrieve the correct structure factor when these effects can not be neglected.

High field amplitudes in the $d=$500 $\mu$m thick grating substrate are responsible for a Kerr-effect induced phase shift $\Delta \phi_{NL} \approx n_2 k_l I d$, with $n_2 = 2.25 \cdot 10^{-16}$ cm$^2$/W. In order to compare to the maximum energy modulation data $\Delta \xi$ plotted in Fig. \ref{fig:gradVfield}, we can integrate Eq. \ref{eq:nonlin} for an on-axis phase-velocity matched particle and obtain the pink dashed curve which does not quite reproduce the observed behavior.

The difference can be explained looking into the additional phase term $\Delta \phi_{PM}$ which takes into account the phase velocity mismatch between the electrons and the accelerating wave inside the DLA. We can identify different contributions to $\Delta \phi_{PM}$ from the effects of pitch, yaw, and the grating-laser combination itself. In fact, in the following, dephasing due to yaw is assumed negligible, since the structure is relatively short and the propagation axis of the e-beam is aligned to the long laser axis. In practice, this angle can not be used to tune dephasing in our setup, since any misalignment in yaw would move the laser off of the grating structure, thus effectively shortening the interaction length. 

We separate the remaining contributions below as
\begin{equation}
    \Delta \phi_{PM}(z,t,\theta) = \Delta \phi (z,t) + \Delta \phi_{IA}(z,\theta)
    \label{Eq:phasemismatch}
\end{equation}
where $\Delta \phi = k_g z - \omega_l t$ is the travelling wave phase and $z$ refers to electron position. If the driving laser wavelength is not matched to the grating period, the effective interaction will be strongly reduced due to the rapidly oscillating integrand in Eq. \ref{eq:nonlin}. The last term $\Delta \phi_{IA}$ is due to the effect of an oblique incidence of the laser onto the structure and can be written as 
\begin{equation}
    \Delta\phi_{IA}(z,\theta) = \frac{\omega_l n}{c} z sin(\theta_0) = \frac{\omega_l}{c} z sin(\theta) 
    \label{eq:varphi}
\end{equation}
where $\theta_0$ is the angle inside the substrate and $n\sin\theta_0 = \sin \theta$ due to Snell's refraction law. Positive $\theta$ corresponds to the laser wave vector, $\vec{k}$, having a positive $z$ component, as seen in Fig. \ref{fig:angle_schematic}.

For an electron beam moving with velocity $v = \beta c$ , we can write the phase change after a grating period as
\begin{equation}
    \Delta\phi_{PM}(\lambda_g,\frac{\lambda_g}{\beta c},\theta) = 2\pi \left[\frac{\lambda_g}{\lambda_l}\left(sin(\theta) - \frac{1}{\beta}\right) + 1\right],
    \label{eq:varphiPeriod}
\end{equation}
where we have set $z = \beta c t$ and $t = \lambda_g /c$.

The phase matching condition is calculated by setting $\Delta \phi_{PM}$ = constant, and can be written as
\begin{equation}
    \label{eq:phaseMatch}
    k_g  - \frac{\omega_l}{c \beta} + \frac{\omega_l}{c} sin(\theta) = 0
\end{equation}
from which it is clear that the illumination angle can be used to compensate for a mismatch in the travelling wave phase, $\Delta \phi$. 
For $\lambda_g = $800 nm and $\lambda_l = $780~nm the resonant condition is satisfied by $\theta = $ 29~mrad for 6 MeV electrons.

The non linear phase due to the laser pulse propagation in the grating substrate and the phase mismatch term jointly combine to determine the phase profile experienced by the particles in the DLA. The black curve in Fig. \ref{fig:gradVfield} is the prediction for an on axis particle, calculated via Eq. \ref{eq:nonlin} for an incident angle 10~mrad away from the phase matching condition. The comparison of experimental data and the predicted curve is optimal for a structure factor of $\kappa = 0.22$. Saturation occurs at an incident field of 6~GV/m and a peak gradient of 0.8~GeV/m in agreement with previous findings \cite{Cesar2018}. These measurements indicate that at high incident intensities in the nonlinear regime, moving away from the phase matching illumination angle results in partial compensation of the Kerr-induced dephasing, thereby raising the incident field at which saturation occurs and switching the concavity of the curve. 

\begin{figure}
    \centering
    \includegraphics[width = 0.45\textwidth]{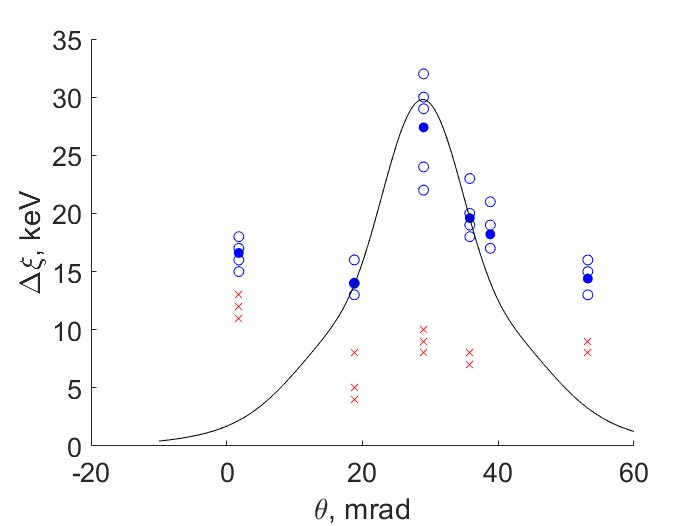}
    \caption{Peak energy modulation as a function of incident angle, with constant incident intensity. The simulated modulation curve in black matches the observed behavior and was obtained assuming an incident laser of 1~mJ and structure factor 0.22, as fit by the gradient scan. In blue is the laser on data, and the red x's denote laser off measurements.}
    \label{fig:angleVmod}
\end{figure}

Finally in Fig. \ref{fig:angleVmod} we plot for a fixed laser intensity the observed peak energy modulation as a function of the angle of incidence $\theta$ of the laser onto the structure. In this plot we overlay the prediction for the energy modulation vs. $\theta$ in black, calculated via Eq. \ref{eq:nonlin}. The width of this curve is inversely proportional to the effective interaction length, $L_{eff} = \sqrt{2} c \tau_l$, and further corroborates $\tau_l = $ 100 fs, as used in the gradient calculations of Fig. \ref{fig:gradVfield}. The addition of the nonlinear phase $\Delta \phi_{NL}(I_0,t)$ to the integrand of Eq. \ref{eq:nonlin} has the effect of broadening the angular acceptance of the grating due to the partial compensation of the nonlinear Kerr phase shift and the dephasing due to the illumination angle. 

In order to frame these results in the longer term plan for DLA development, it is important to calculate in the grating reference frame the phase advance between adjacent periods by setting $z = \lambda_g$ in Eq. \ref{eq:phaseMatch}.
\begin{equation}
    \Delta\phi_{IA}(\lambda_g,\theta) = 2 \pi \frac{\lambda_g}{\lambda_l} sin(\theta).
\end{equation}

\begin{figure}
    \centering
    \includegraphics[width = 0.45\textwidth]{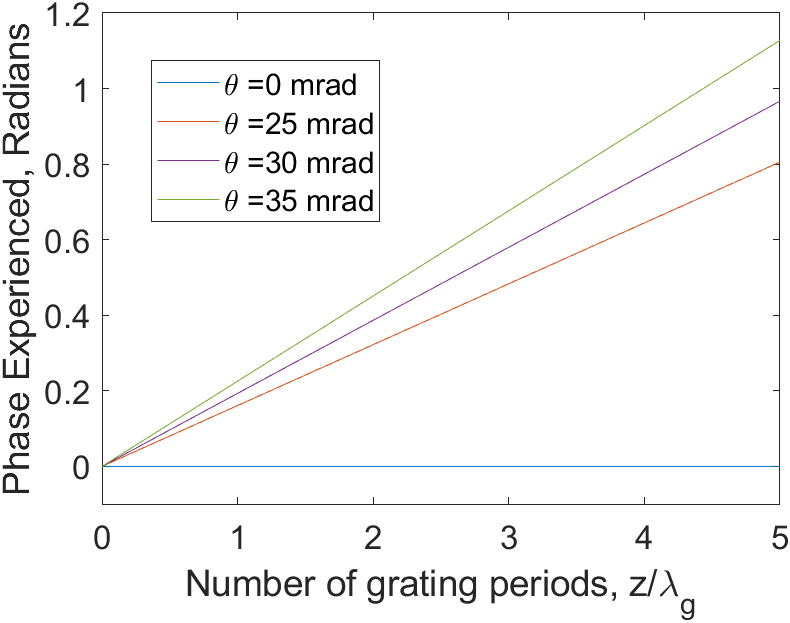}
    \caption{Plot of the phase advance within the grating structure due to the angular offset of the incident laser. The observation of DLA acceleration signal up to 30 mrad incidence angle, corresponding to a phase advance of 194 mrad/period, confirms the large acceptance bandwidth of the double grating structure.}
    \label{fig:bandwidth}
\end{figure}

This quantity is relevant for future planned experiments to modulate the incident laser phase, increasing the interaction length to the cm scale by taking advantage of ponderomotive focusing effects. In the various schemes put forward \cite{cesar:alloptical, Ody:SHARD}, the predicted maximum phase advance per period is 6.5 mrad; in this experiment, we demonstrated that the double-grating DLA accelerator still works efficiently for phase shifts of over 200 mrad per period, as seen is Fig. \ref{fig:bandwidth}. This result demonstrates a flexibility in grating parameter choices, showing that it is possible to accelerate electrons by DLA provided that the driving phase is controlled to ensure resonant interaction. This flexibility could enable the use of structures that are not purposely fabricated for specific DLA setups. It also allows for relaxed tolerances in the structure dimensions, as shaping the excitation pulse can correct for errors.

\section{Conclusion}
In this paper we present first experimental results of acceleration in a DLA structure driven by a laser-wavelength mismatched with the grating period. The angle of incidence is used to compensate for the mismatch and recover resonant interaction with 6 MeV electrons. 

These results highlight the interplay between nonlinear effects and angle-of-incidence phase velocity mismatch in determining the induced energy modulation. Detailed control of the laser phase experienced by the electrons in the structure is critical to maximize the interaction length and induced energy modulation.

The results provide an important validation for the wide phase acceptance of double grating DLA structures. Such structures will be exploited in future experiments where the driving phase will be rapidly varied to ponderomotively focus the e-beam inside the DLA, thereby increasing the interaction length. 

This work has been supported by the ACHIP grant from the Gordon and Betty Moore Foundation (GBMF4744) and by U.S. Department of Energy grant DE-AC02-76SF00515. This material is based upon work supported by the National Science Foundation Graduate Research Fellowship Program under Grant No. DGE-1650604. Any opinions, findings,and conclusions or recommendations expressed in this material are those of the author(s) and do not necessarily reflect the views of the National Science Foundation.

\end{document}